\documentclass[aps, preprint, 12pt]{article}
\usepackage{times}
\usepackage{amssymb}
\usepackage{amsmath}
\usepackage{array}
\usepackage[margin=2.5cm]{geometry}
\usepackage{fancyhdr}
\usepackage{cite}
\usepackage[format=hang,=small,labelfont=bf]{caption}
\usepackage{authblk}

\usepackage{color}
\usepackage{pdfcolmk} 
\usepackage{setspace}
\usepackage{soul}
\usepackage[hidelinks]{hyperref}

\ifx\pdfoutput\undefined
\usepackage{graphicx}
\else
\usepackage[pdftex]{graphicx}
\usepackage{epstopdf}
\fi

\usepackage{pdflscape} 

\usepackage[version=3]{mhchem} 

\newcolumntype{x}[1]{>{\raggedright\hspace{0pt}}p{#1}}

\graphicspath{{.},{./f/}}

\title{Machine learning accelerated discovery of corrosion-resistant high-entropy alloys}

\author{Cheng Zeng$^{\text{*}, \ddag}$, Andrew Neils, Jack Lesko, Nathan Post$^{\text{*}}$}
\affil{The Roux Institute, Northeastern University, Portland, Maine, 04101, United States. \\ $^\ddag$ The Experiential AI Institute, Northeastern University, Boston, Massachusetts, 02115, United States \\$\text{*}$ Corresponding authors: Email: c.zeng@northeastern.edu and n.post@northeastern.edu, Tel: +1 401-396-6668 and +1 781-605-8671\\}

\begin{document}
\maketitle


\begin{abstract}
\noindent
Corrosion has a wide impact on society, causing catastrophic damage to structurally engineered components. An emerging class of corrosion-resistant materials are high-entropy alloys. However, high-entropy alloys live in high-dimensional composition and configuration space, making materials designs via experimental trial-and-error or brute-force ab initio calculations almost impossible. Here we develop a physics-informed machine-learning framework to identify corrosion-resistant high-entropy alloys. Three metrics are used to evaluate the corrosion resistance, including single-phase formability, surface energy and the compactness of oxide films formed on an alloy surface evaluated by Pilling-Bedworth ratios. We used random forest models to predict the single-phase formability, trained on an experimental dataset. Machine learning inter-atomic potentials were employed to calculate surface energies and Pilling-Bedworth ratios, which are trained on first-principles data fast sampled using embedded atom models. A combination of random forest models and high-fidelity machine learning potentials represents the first of its kind to relate chemical compositions to corrosion resistance of high-entropy alloys, paving the way for automatic design of materials with superior corrosion protection. This framework was demonstrated on AlCrFeCoNi high-entropy alloys and we identified composition regions with high corrosion resistance from a wide range of compositions. Machine learning predicted lattice constants and surface energies are consistent with values by first-principles calculations. The predicted single-phase formability and corrosion-resistant compositions of AlCrFeCoNi agree well with experiments. This framework provides a computationally efficient approach to navigate high-dimensional composition space of high-entropy alloys. It is general in its application and applicable to other complex materials, enabling high-throughput screening of material candidates and potentially accelerating the iteration of integrated computational materials engineering.
\end{abstract}

\newpage

{\singlespace \footnotesize \noindent \textbf{Keywords:} High-entropy alloy, Corrosion protection, Machine learning potential, Random forest classification}

Graphical abstract:

\begin{center}\includegraphics[width=3.3in]{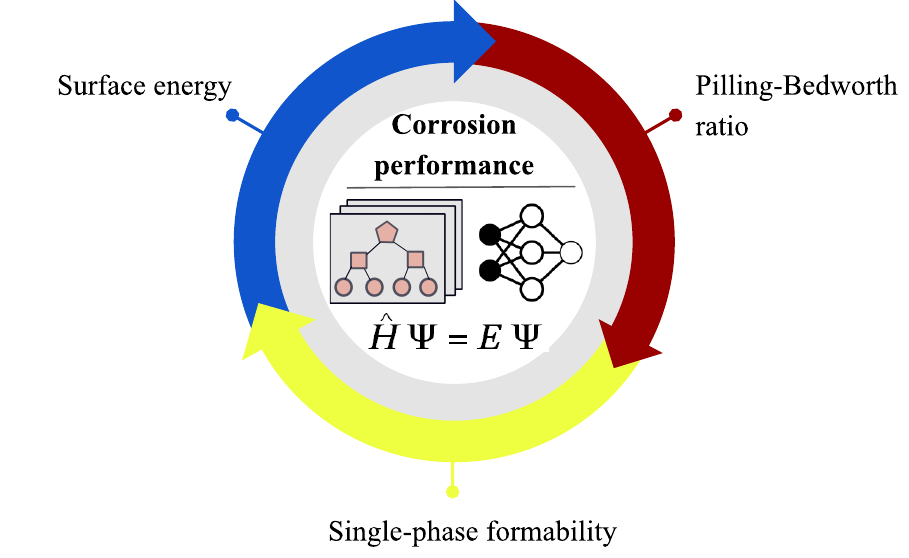}\end{center}

\newpage

\section{Introduction}

High-entropy alloys are generally defined as alloys comprising no less than four elements and the percentage of each principal element is between 5 at.\% and 35 at.\%.
The high-entropy concept was coined by Cantor~\cite{cantor2004} and Yeh~\cite{yeh2004} for equiatomic alloys with no less than five elements in 2004 almost the same time.
The definition has been slightly extended to non-equimolar alloys with no less than four principal elements.
This new class of materials has attracted increasing attention, found to display superior materials performance for mechanical properties~\cite{gludovatz2014,yang2018,li2016, ding2021, wang2012}, radiation resistance~\cite{kumar2016,lu2019} and corrosion resistance~\cite{shi2018, qiu2015, fu2021}.
The high entropy of mixing usually leads to the formation of a disordered single phase for high-entropy alloys, such as face-centered cubic (FCC), body-centered cubic (BCC) and hexagonal closely-packed structures (HCP)~\cite{steurer2020,zhao2016}.
The homogeneous single phase improves passivity.
In addition, high-entropy alloys can consist of elements with high passivation potency such as nickel, chromium, aluminum and titanium, leading to high pitting corrosion resistance.

Conventional corrosion-resistant alloys are mostly found by serendipity.
Advances in physical theories, computational hardware and algorithms allow for rapid screening of candidate materials, paving the way for integrated computational materials engineering which aims to demystify the linkage between process, structure, property and performance.
However, computational screening of corrosion-resistant alloys is challenging in that many factors can influence corrosion performance, including environmental conditions, chemical compositions and microstructures.
Moreover, fundamental understanding of corrosion and various corrosion types adds more complexity to the material design.
Recent works have been focused on building reliable databases for corrosion informatics, identifying reliable descriptors for corrosion performance and understanding the corrosion kinetics with multi-physics simulations~\cite{taylor2018, nyby2021, ansari2021}.
Nyby et al. compiled a database for four types of alloys with an emphasis on six metrics used to describe their localized pitting corrosion~\cite{nyby2021}.
Diao et al. collected a dataset for low-alloy steel and built machine learning models to predict their corrosion rate~\cite{diao2021}.
Roy et al. used machine learning algorithms to select the top three descriptors for prediction of the corrosion rates, including pH of the medium, halide concentration and composition of elements with the minimum reduction potential~\cite{roy2022}.
Taylor et al. identified a number of corrosion descriptors, such as cohesive energies, oxide formation energies and surface enrichment of passive elements, and related those descriptors to corrosion resistance with respect to surface passivation, dissolution and microstructure control~\cite{taylor2018}.
Ke and Taylor reviewed the role of density functional theory (DFT) in modeling corrosion, and they pointed out corrosion metrics accessible by DFT, including oxygen and chloride adsorption energy, dissolution potential and surface energy~\cite{ke2019}.
Other computational methods based on peridynamics and phase-field modeling are often used to study the evolution of pitting corrosion~\cite{ansari2021, chen2021}.

Unfortunately, the complexity of corrosion process makes it almost impossible to relate chemical compositions and microstructures of alloys directly to the corrosion performance.
The vast composition and microstructure space of high-entropy alloys create complexity for the materials design problem.
A workaround is multi-objective optimization based on empirical rules, which allows for screening material candidates with relative superior corrosion resistance.
While some data-driven approaches and first-principles calculations exist to identify corrosion descriptors, those data-driven methods in nature lack physical insights and first-principles calculations are costly computationally.
A physically meaningful and efficient approach to relating compositions with corrosion performance is still lacking.

The objective of this work is to bridge the technical gap for locating high-entropy alloys with potential high corrosion resistance in the high-dimensional composition space, in particular for pitting corrosion.
We focused on pitting corrosion because the rate of localized pitting corrosion can be faster than uniform corrosion by orders of magnitude, hence pitting corrosion is more critical in applications where it exists~\cite{nyby2021}.
Pitting corrosion normally occurs under non-equilibrium conditions and the pitting event may be subject to inherent randomness, making it almost impossible to be predicted exactly~\cite{coelho2023, wu1997}.
A complete picture of pitting corrosion requires multi-scale and multi-physics simulations together with stochastic modeling to understand the formation of passive film, passive film breakdown and pit growth stability~\cite{li2021}, which to the best of our knowledge are not available.
Pitting corrosion resistance is empirically associated with the ability of alloys to form a passive film, protectiveness of the passive film and pitting growth rate when the passive film breaks down.
Although it is challenging to predict the exact pitting process, it is well-acknowledged that a more compact passive film and a slower pit growth rate make the structures less susceptible to pitting corrosion~\cite{li2021}.
In this work, we chose three corrosion metrics considered to be influential to pitting corrosion, including \textit{single phase formability}, \textit{Pilling-Bedworth ratio} of passive elements, \textit{surface energy}.
All three metrics used here are either rationalized in theory or validated in experiments.
Finding single-phase solid solutions is a classic strategy for the design of corrosion-resistant alloys since single-phase alloys tend to form more homogeneous and compact passive films~\cite{fu2021}.
Pilling-Bedworth ratio is a well-acknowledged metric for evaluation of the compactness of oxide films formed on an alloy surface as it accounts for the volume change when a passive film forms on a metal surface~\cite{xu2000, tan2016}.
Surface energies were first used by Song  et al to describe the experimental relationship between surface orientation and corrosion resistance, and it was found that a more closely packed surface with a lower surface energy tend to exhibit superior corrosion resistance~\cite{song2010}.
In addition, surface energies have often been used in atomistic simulations to gain insights into corrosion performance of alloys~\cite{taylor2018,ke2019}.
Moreover, one may also need to consider the key role of chloride ions in pitting corrosion.
A more comprehensive optimization of corrosion-resistant high-entropy alloys may directly model the chloride effects via simulating the adsorption of chloride on a metal surface or on a passive film surface, which can affect the breakdown of passive films or the fast growth of the initiated pit~\cite{bouzoubaa2009, taylor2018, ke2019}.
More sophisticated models study the fast diffusion of chloride ions through the defects in the passive films and the accumulation of those ions at the interface between metal and oxides~\cite{lin1981, zhang2018}.
Despite not modeling the chloride effect explicitly in this work, it is arguable that single-phase structures are more corrosion resistant in a salt solution with chloride ions because single phase structures tend to form a homogeneous passive film which will provide fewer fast diffusion channels compared to a heterogeneous passive film.
Besides, lower surface energies are more likely to show a weaker binding between metal and chloride ions at the metal/oxide interface when a pit is initialized~\cite{taylor2018}.
A physics-informed machine learning (ML) framework was then introduced to quantify the three corrosion metrics for a wide range of compositions of high-entropy alloys.
Details of theories and how to calculate the three corrosion metrics will be elaborated in the subsequent section.
We tested this framework for AlCrFeCoNi high-entropy alloys by studying the three corrosion metrics as a function of compositions of high-entropy alloys, based on which compositions with high corrosion resistance will be identified and compared to experiments.
AlCrFeCoNi high-entropy alloys were considered because they belong to an emerging class of materials with superior mechanical properties and corrosion resistance~\cite{ren2022, shi2018}.

\section{Theories and Methods}

The aim of this work is to develop machine learning methods to quantify three corrosion metrics for high-entropy alloys.
There are two types of machine learning models involved in the prediction of the three metrics.
One is a random forest model trained on an experimental dataset where the inputs are the chemical formulas and the output is a boolean to show whether a single-phase structure is formed or not.
The other machine learning model is a machine learning potential trained on first-principles data where the inputs are atomic structures and the outputs are potential energy and forces of the structures obtained by first-principles calculations.
The single-phase formability is calculated by the random forest model, whereas Pilling-Bedworth ratios and surface energies are computed by the machine learning potential.
The overall workflow of how to train these two types of machine learning models is illustrated in Figure~\ref{fig:workflow}.
In this section, we discuss the concept and training of a state-of-the-art machine learning potential as well as how to quantify each of the corrosion metrics by the trained machine learning models.

\subsection{Machine learning potentials}
Potential energy surfaces (PESs) represent one-to-one mappings between atomic positions ($\{R\}$) and potential energy ($E$) of a material system.
PESs provide a plethora of information for material systems.
For example, local minima on PESs represent stable states and the minimum energy trajectory connecting two local minima indicates a fundamental reaction pathway.
The most often used methods to build reliable PESs are DFT calculations.
However, standard DFT calculations are limited to hundreds of atoms due to the formidable $\mathcal{O}(M^{2-3})$ scaling with system sizes (M), such as numbers of basis sets, atoms or electrons~\cite{bowler2012}.
It is thus computationally prohibitive to sample all points on \textit{ab initio} PESs.
One should note that DFT, first-principles and \textit{ab initio} calculations are used interchangeably as they have the same meaning in this work.
In the past decade, fitting \textit{ab initio} PESs with machine learning (ML) algorithms  have gained increasing momentum, and the ML-fitted PESs are termed machine learning potentials (MLP).
Most MLPs relies on the nearsightedness principles~\cite{kohn1996}, also known as \textit{all chemistry is local}, implying that the total potential energy of a system with $N$ atoms can be largely decomposed into a linear sum of all atomic contributions and each atomic contribution comes from the atom $i$ interacting with neighboring atoms in a cutoff region, written as Eq.~\ref{eq:locality}.

\begin{equation}\label{eq:locality}
E = E(\{R\}) = \sum_{i=1}^N E_i = \sum_{i=1}^N E_i^{\mathrm{(local)}}
\end{equation}

\begin{figure}
\centering
\includegraphics[width=6.5in]{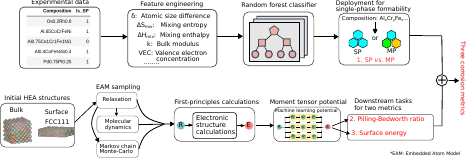}
\caption{Workflow of machine learning accelerated discovery of corrosion-resistant high-entropy alloys.}
\label{fig:workflow}
\end{figure}

\noindent
Thanks to and only because of the nearsightedness principle, MLPs can be trained with small-size first-principles data while allowing for reliable predictions on much larger systems~\cite{behler2007}.
It should be noted that the nearsightedness of first-principles calculations and machine learning algorithms should be well aligned to strike a good balance between computational efficiency and prediction accuracy~\cite{zeng2022}.
A variety of ML algorithms have proven effective in fitting \textit{ab initio} PESs, such as neural networks~\cite{behler2007,xie2018}, Gaussian process~\cite{bartok2010} and kernel ridge regression~\cite{rupp2012}.
MLPs find applications in many fields, ranging from small molecules, to nanoparticle alloy catalysts and extended systems~\cite{batzner2022,zeng2023,behler2021}.
In this work, we employed a class of machine learning potentials termed moment tensor potentials (MTPs) for the high-entropy alloys which were found to be superior to other types of MLPs when tested on single-element systems by various simulation tasks~\cite{zuo2020}.
Readers should refer to the work of Shapeev for implementation details of MTPs~\cite{shapeev2016,novikov2020}.
MTPs were trained with systematically generated and first-principles calculated training data for high-entropy alloys AlCrFeCoNi, as outlined in the bottom row of Figure~\ref{fig:workflow}.
We primed the algorithm with FCC bulk and surface structures.
For each of the initial structures, atomic positions and lattice geometry are both optimized to find the stable structure, the process of which is termed structure optimization, also known as relaxation.
The structure optimization used the embedded atom method (EAM) developed by Farkas and Caro~\cite{farkas2020}.
Starting with relaxed structures and using EAM, we also sampled a diverse pool of atomic configurations via molecular dynamics and Monte-Carlo simulations.
The molecular dynamics simulations were used to perturb atomic positions, whereas the Monte-Carlo simulations were adopted to simulate exchange of two different atoms.
Electronic structure calculations were performed with an open Python package Grid-based Projected Augmented Waves (GPAW)~\cite{dohn2017} to refine the energy and forces for a part of the EAM-sampled configurations.
Additionally, we carried out first-principles calculations for simple bulk and surface structures with numbers of elements ranging from one to five.
Simple bulk structures included special quasi-random structures generated using the tool in the alloy theoretic automated toolkit (ATAT) for simple bulk structures with more than two elements to best approximate a random solid solution~\cite{zunger1990,van2013}.
In total, 1569 first-principles structures were curated.
We performed a 5-fold cross-validation on the generated structures.
We then trained a MTP upon the full dataset, and we used the MTP to carry out simulations needed to calculate PBRs for the oxidation of Cr and surface energies of FCC(111) facets.
Atomic structures were created and manipulated with the Atomic Simulation Environment (ASE)~\cite{larsen2017} and LAMMPS~\cite{thompson2022}.
Computational settings of MTPs and GPAW calculations and details of the training data can be found in the SI.
MTP enabled simulations to calculate relevant corrosion metrics are also elaborated in the SI.
Scripts and notebooks for atomistic modeling and curation of training data will be supplied as a supporting dataset.

\subsection{Single phase formability}
It is crucial to form a homogeneous single phase for enhanced corrosion protection because it enhances passivity and prevents the fast galvanic corrosion.
A physically rigorous approach to model single phase formation is thermodynamic modeling carried out with CALPHAD.
The reliability of CALPHAD calculations is determined by the quality of experimental data as well as relevant first-principles calculations~\cite{mao2017}.
Instead of thermodynamic modeling, we used random forest models trained on an experimental dataset to predict the probability of forming single phases for an arbitrary alloy composition.
The workflow for single phase formability is shown in the top row of Figure~\ref{fig:workflow}.
The experimental dataset was summarized by Yan et al.~\cite{yan2021}.
The raw data in total has 1807 entries and it takes input as the chemical compositions and output as the indicator for single phase formability.
Single-phase alloys are labeled as '1', whereas multiple-phase alloys are labeled as '0'.
It should be noted that phase formations are also dependent on the manufacturing processes and thermal history.
It is assumed that most alloys found in this experimental dataset were processed with similar techniques and environmental conditions, and the remaining exceptions represent outliers and noise in the dataset, whose impact on the robustness of ML models will be diminished by a cross-validation strategy due to the averaging effect.
Each input composition was converted to eight physical descriptors including atomic size difference, mixing enthalpy, mixing entropy, Pauli electronegativity difference, molar volume, bulk modulus, melting temperature and valence electron concentration, as considered by Yan et al~\cite{yan2021}.
Next, random forest models were trained to relate the eight descriptors to the single phase formability.
The trained random forest models are thus able to predict if alloys will form single phase or multiple phases for a given composition.
For the purpose of model validation, we held out 20\% of the entire dataset for testing, which were used to examine the prediction accuracy of the final model on some unseen dataset, hence avoiding overfitting.
Five-fold cross-validation was used on the remaining 80\% dataset to tune the hyperparameters and to train the random forest classifier.
The entire dataset was shuffled before splitting with a given random state to a test set and a data set for cross-validation, and ten random states were used to estimate uncertainties due to data splitting.
One should however note that the model trained in this work may differ from that by Yan et al as the constructed eight descriptors may be different, and Yan et al only used five descriptors rather than all the eight descriptors used in this work.
We attempted to reproduce Yan's results by matching available model settings, and we found a prediction accuracy of 88\% on the test set, less accurate than that using all eight features.
The code to reproduce our computational experiment is included in a public github repository.
Details of each descriptor, exploratory data analysis (Figure S1-S3), and hyperparameters of random forest models are included in the supporting information (SI).

\subsection{Pilling-Bedworth ratio}
Pilling-Bedworth ratio (PBR) was used to describe the compactness and growth stress of an oxidation process.
It describes the volume change due to oxidation on an alloy surface, which follows Eq.~\eqref{eq:pbr} with respect to the oxidation of a metallic element B.

\begin{equation} \label{eq:pbr}
PBR_{\textrm{B}} = \frac{\mathrm{Volume~of~a~mole~of~B_xO_y}}{\mathrm{Volume~of~x~moles~of~B~in~metal}}
\end{equation}

It is well accepted that when $PBR < 1$, the formed oxide offers no protection to the alloy surface.
If $1 \leq PBR \leq 2$, the oxide forms a passive layer and prevents structural alloys from direct corrosion although some compression stresses develop inside the oxide.
When $PBR \gg 2 $, the compression stresses become significant, causing the breakdown of the oxides.
This simple analysis well explains that corrosion-resistant alloys typically contain Al, Zr, Ni, Ti, Fe or Cr whose PBR values are larger than 1 and not much larger than 2~\cite{zeng2016,fu2021,zeng2017}.

When it comes to the oxidation of alloys, one or more elements may oxidize and form passive layers.
Hence we need to identify elements that are thermodynamically preferential for oxidation.
We can then calculate the PBR of the identified passive element by analyzing the volume change due to its oxidation.
Xu and Gao introduced methods to compute PBR for the oxidation of alloys~\cite{xu2000}.
There are two possible cases for PBR values of alloys, depending on the relative diffusion rate of the passive element in alloys versus that in oxides.
Generally the diffusion rate of passive elements within alloys are much  faster relative to the rate within oxides so that alloy compositions near the surface can maintain a stoichiometry close to the original composition.
For example, the diffusion coefficient of Cr in CrCrFeMnNi high-entropy alloys at 900 $^\circ$C is about $10^{-12}$ cm$^2$/s while self-diffusion of Cr in \ce{Cr2O3} has a coefficient on the order of $10^{-21}$ to $10^{-17}$ cm$^2$/s~\cite{sabioni2005, tsai2013}.
We recommend that readers consult the work of Xu and Gao~\cite{xu2000} for calculation details of PBR for oxidation of alloys.
For our benchmark material system AlCrFeCoNi, we examined the passivation of the Cr element.
We chose to study Cr oxides because experiments by Shi et al indicates that it is the more compact Cr oxides that offer the major protection against pitting corrosion than the more porous Al oxides/hydroxides when it comes to the passivation of AlCrFeCoNi~\cite{shi2020} although thermodynamic data favors the formation of Al oxides over Cr oxides, as tabulated in Table S1 of SI.
Therefore, we consider the oxidation of Cr, which forms \ce{Cr2O3} with a mole weight of 102 g/mol and a density of 5.22 g/cm$^3$.
The volume of Cr in the alloy was calculated using a FCC crystal whose lattice parameters were obtained by machine learning potentials.

\subsection{Surface energy}
Surface energies are calculated by the trained moment tensor potential.
The electrochemical dissolution rate $I_A$ of a metal `A' with an exposed crystal plane (h,k,l) at a temperature $T$ follows the relation:

\begin{equation}\label{eq:rate}
I_{A,(h,k,l)} \propto \exp \left(\frac{\alpha  \gamma_{(h,k,l)}}{R T}\right)
\end{equation}

\noindent
where R is the gas constant, $\gamma_{(h,k,l)}$ is the surface energy and $\alpha$ is a transition coefficient to relate surface energy to dissolution activation energy.
Ramachandran and Nosonovsky found that a lower surface energy leads to a more hydrophobic surface, and hydrophobic surfaces tend to show higher corrosion resistance~\cite{ramachandran2015}.
In this work, we used surface energy as a metric to describe the trend of average dissolution of atoms on the crystallographic plane FCC(111) of AlCrFeCoNi alloys with different compositions.
An FCC(111) facet wast used because of its high stability over other types of facets.
It is arguable that a higher surface energy is associated with a higher average dissolution rate, resulting in faster pitting growth, although a more rigorous treatment may need to take into account sequential atom-by-atom dissolution on a metallic surface constrained by broken passive films formed on top of it, which we elected not to consider for the sake of simplicity.
Surface energy of a facet reads as:

\begin{equation}\label{eq:gamma}
\gamma = \frac{E_{\mathrm{slab}} - E_{\mathrm{bulk}}}{2 A}
\end{equation}

\noindent
where $E_{\mathrm{slab}}$ and $E_{\mathrm{bulk}}$ are respective potential energies of the FCC(111) facet and bulk cell, and A is the exposed area of the facet.
The bulk cells used to calculate surface energies are of L1$_2$ structures, and the surface structures are the putative most stable structures found by Markov chain Monte-Carlo (MCMC) simulations.
All terms in Eq.~\ref{eq:gamma} were found by atomistic modeling using machine learning potentials.
The details of MCMC simulations are provided in the SI.

\subsection{Mapping corrosion metrics with respect to Al and Cr compositions in AlCrFeCoNi}
We tested the above methods on predicting the three corrosion metrics for AlCrFeCoNi high-entropy alloys. We varied the compositions of Al and Cr while equalizing remaining Fe, Co and Ni compositions.
For a given composition, its single-phase formability was calculated by a random forest classifier and its Pilling-Bedworth ratio and surface energy were quantified by the MTPs.
Therefore, we mapped the three corrosion metrics as a function of AlCrFeCoNi compositions, based on which we can identify composition regions with desired values for all corrosion metrics, which are potentially associated with superior relative corrosion resistance.
We changed Al compositions in the range of 0--25 at.\%, and Cr compositions in the range of 10--30 at.\% (see Figure~\ref{fig:results}).
The lower bound for the Cr composition was set as 10\% because Cr is the passive element, and a percolation model for passivation of alloys suggests that the smallest amount of elements to enable passivation is around 10\%~\cite{sieradzki1986}.
In other words, an alloy only forms a continuous and protective passive film with the passive element being of no less than 10 at.\%.
For single-phase formability, an interval of 1\% was used for both Al and Cr composition mesh grids as the inference by the trained random forest classifier for each composition took less than 1 second.
For PBR$_{\rm{Cr}}$, an interval of 5\% was used to find the lattice parameters of L1$_2$ bulk cells using MTPs.
The lattice parameters of MTP-relaxed structures were then fitted by a linear regression as a function of Al and Cr compositions.
The fitted function was used to calculate lattice parameters for arbitrary Al and Cr compositions.
The volume of Cr will be used to calculate PBR$_{\rm{Cr}}$.
In terms of surface energies, a composition interval of 5\% was used to generate structures needed.

\section{Results and discussion}

\subsection{ML prediction accuracy and transferability}

ML models are often criticized for their poor transferability to data that are not existent in the training data set.
While issues with ML transferability can be mitigated via a transfer learning approach where models are fine tuned to assimilate new data~\cite{vangrunderbeek2023}, limitations in transferability often originate from inaccessibility to a wide range of data and the way that the data are represented and encoded in the machine learning models.
As a result, it is of great importance to evaluate ML model performance before we deploy the models.

\subsubsection{Random forest classifier}
The experimental dataset used to train the random forest classifier includes in total 1807 entries.
The 1807 data points were split to 80\% and 20\% for cross-validation and test, respectively.
Hence 361 data points were used for testing, and ten different random states for the data splitting were used to obtain the standard deviation of model prediction accuracy on the test set.
The random forest classifier gave a prediction accuracy of 89\% on the test set with a standard deviation of 1\%.
The best model with the highest accuracy on the test set was chosen for subsequent inferences.
We also studied the feature importance using shapley values based on game theories~\cite{merrick2020}.
Mixing entropy, atomic size difference and melting temperature were identified as the top three most important features, largely consistent with the work of Yan et al~\cite{yan2021}.
More feature importance results can be found in Figure S4 of SI.

\subsubsection{Moment tensor potentials}
For the 5-fold cross-validation over the full 1569 structures, we observed 6.2 $\pm$ 1.6 meV/atom and 0.086 $\pm$ 0.016 eV/\AA~ for respective energy and force average absolute differences (numbers following `$\pm$' represent the standard deviations across different folds).
The MTP trained on the full dataset gave $\sim$5 meV/atom for the average absolute difference of energy  and 0.058 eV/\AA~for the average absolute difference of atomic forces.
To further validate the MTPs, we compared predicted lattice constants of single-element FCC crystals to values by DFT.
We also compared the predicted surface energies of single-element FCC(111) facets with DFT.
The comparison is summarized in Table.~\ref{tab:mtp-val}.
One can see that MTP-predicted lattice constants are close to DFT calculations, with relative deviations around 1\%.
In terms of FCC(111) surface energies, although large deviations exist for elements Ni, Co and Al, the relative order of surface energy magnitude by MTP is in accordance with that by DFT.

\begin{table}
\small
\centering
 \caption{Lattice constants and FCC(111) surface energies for single-element structures: DFT \textit{versus} MTP.}
\begin{tabular}{ccc|cc}
\hline \hline
 {} & \multicolumn{2}{c}{Lattice constant [\AA]} & \multicolumn{2}{|c}{Surface energy [J/m$^2$]} \\
\hline
Element & MTP & DFT & MTP & DFT \\
\hline
Al & 4.08 & 4.04 & 0.77 & 0.86 \\
Cr & 3.62 & 3.62 & 2.61 & 2.65 \\
Fe & 3.44 & 3.46 & 2.49 & 2.45 \\
Co & 3.49 & 3.46 & 1.87 & 2.12 \\
Ni & 3.51 & 3.52 & 1.93 & 2.14 \\
\hline \hline
\end{tabular}
\label{tab:mtp-val}
\end{table}

\begin{figure}
\centering
\includegraphics[width=4.0in]{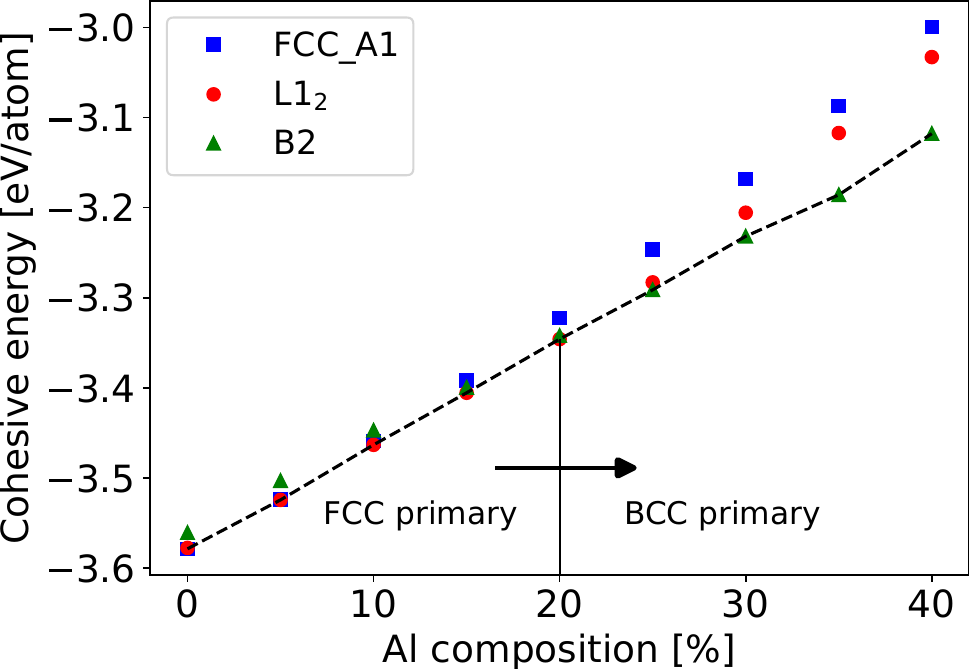}
\caption{Cohesive energies of the FCC\_A1, L1$_2$, and the ordered B2 phases for Al$_x$(CrFeCoNi)$_{100-x}$ as a function of Al compositions. The most stable phases at all Al compositions are connected with dashed lines to guide the eyes.}
\label{fig:alx-ps}
\end{figure}

We also compared the phase stability among various single-crystal structures, including FCC random alloys (FCC\_A1), FCC L1$_2$ ordered structures and BCC B2 structures.
This comparison was used to test the ability of MTPs to predict the most stable phase of Al$_x$(CrFeNiCo)$_{100-x}$ as a function Al compositions.
Experimental observation and EAM-based calculations suggest that a low Al composition favors FCC-type phases while B2 phases are thermodynamically more stable at higher Al compositions~\cite{wang2012,farkas2020}.
We calculated the cohesive energies of L1$_2$ and B2 for Al compositions up to 40\%, with all Al in one sublattice and the remaining four elements randomly distributed.
The FCC\_A1 structures were generated by randomly placing the atoms in a FCC lattice.
Figure~\ref{fig:alx-ps} shows that at low Al contents (0--10\%), L1$_2$ and FCC\_A1 are both more stable than B2 phases.
When Al compositions increase (10--20\%), the ordered L1$_2$ becomes the most stable phase.
In comparison, larger Al compositions ($> 20\%$) favor the formation of ordered B2 phases, in good agreement with well-parameterized empirical potentials and experiments~\cite{farkas2020,wang2012}.
The dashed line represents the most stable phases at each Al composition.
One should note that the first-principles data used to train MTPs only consist of FCC structures.
Despite not seeing any BCC structures, the MTPs accurately predicted the trends of phase stability that are consistent with experiments and first-principles data, indicating decent transferability of the MTPs.

\subsection{Variation of three corrosion metrics as a function of Al and Cr compositions in AlCrFeCoNi}

We studied the three corrosion metrics as a function of Al and Cr compositions for AlCrFeCoNi high-entropy alloys.
This specific high-entropy system was examined because of its superior corrosion performance and the availability of extensive experimental corrosion data, which can be compared to the ML predictions~\cite{shi2018, shi2021, shi2022}.

\begin{figure}
\centering
\includegraphics[width=6.0in]{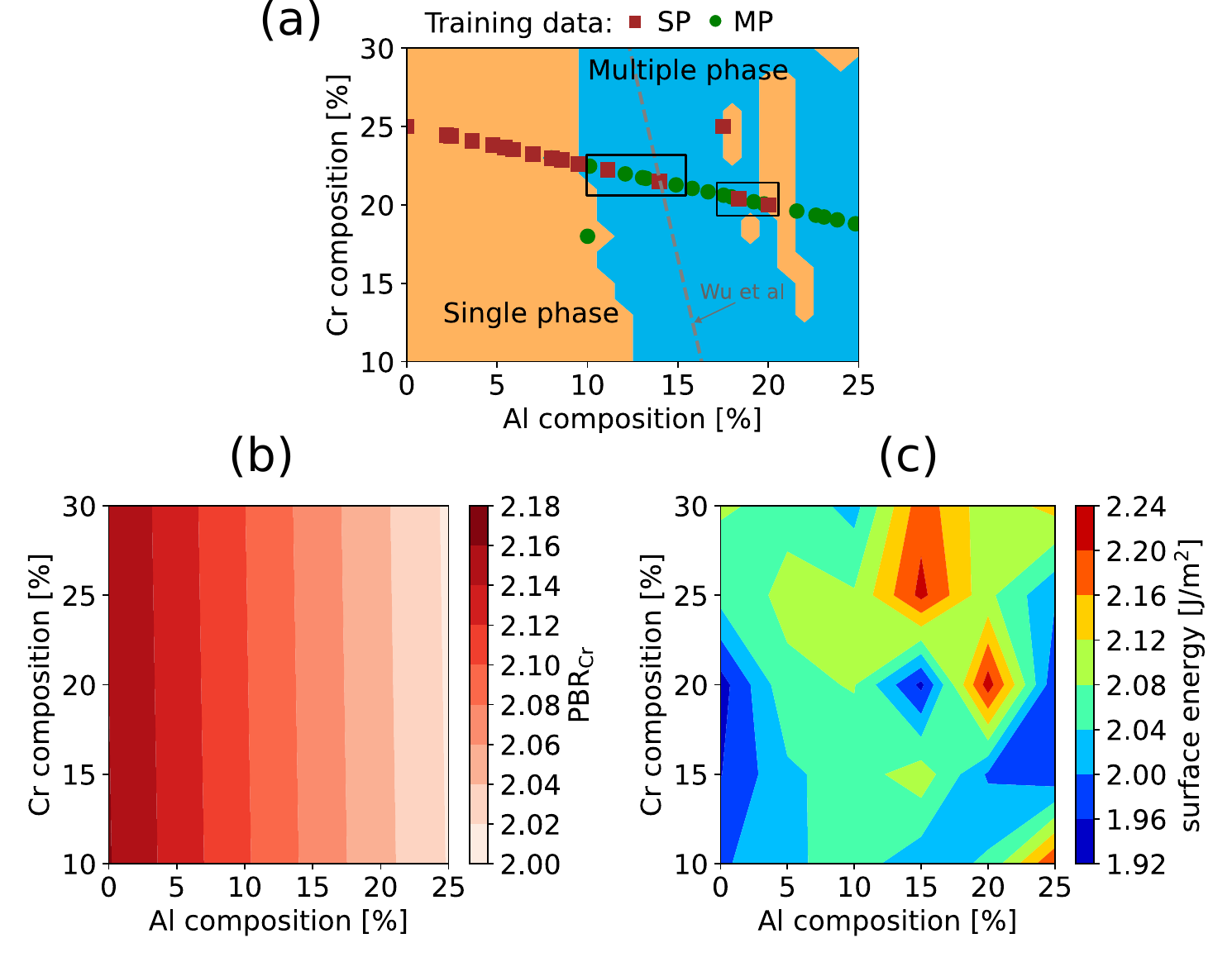}
\caption{Single phase formability (a), Pilling-Bedworth ratios for oxidation of Cr (b) and FCC(111) surface energies (c) as a function of Al and Cr compositions in Al$_x$Cr$_y$(FeCoNi)$_{100-x-y}$. The training data involving AlCrFeCoNi high-entropy alloys are included as scatter points in (a). Red squares and green circles in (a) represent single phase and multiple phase data, respectively. The grey dashed line in (a) is roughly the decision boundary by Wu et al~\cite{wu2020}. The regions where multi phases and single phase are almost overlapped are marked with black boxes in (a).}
\label{fig:results}
\end{figure}

\begin{figure}
\centering
\includegraphics[width=4in]{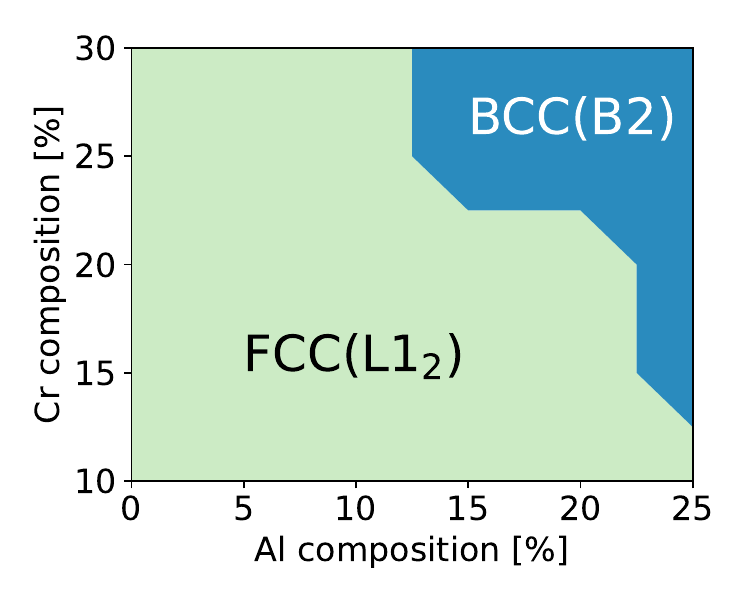}
\caption{Single crystal phase stability (L1$_2$ versus B2) as a function of Al and Cr compositions.}
\label{fig:phase-stability}
\end{figure}

Single phase formability, Pilling-Bedworth ratios and surface energies are shown in respective Figure~\ref{fig:results} (a), (b) and (c) as a function of Al and Cr compositions in AlCrFeCoNi.
The random forest classifier used to predict single phase formability gave a prediction accuracy of 91\% on the training data consisting of AlCrFeCoNi, although there exists some noise as both multiple-phase and single-phase AlCrFeCoNi structures  are found in small composition regions marked with black boxes.
In Figure~\ref{fig:results}(a), one can locate a decision boundary at around 10\% Al composition to separate single-phase and multiple-phase alloys.
The decision boundary is mostly determined by Al compositions and slightly associated with Cr compositions.
An increase of Cr compositions marginally shifts the boundary to a lower Al composition.
This decision boundary agrees well with the experimental results by Wu et al~\cite{wu2020}, which are indicated by the grey line in Figure~\ref{fig:results}(a).
Nevertheless, one can note that predictions of single-phase formation at high Al composition regions are confounded by the noisy training data, implying high prediction uncertainties in this region.
In terms of tendency to form corrosion-resistant passive films, it is also crucial to form specific single-phase structures.
It was experimentally found that FCC AlCrFeCoNi crystals tend to be more resistant to pitting corrosion than BCC counterparts~\cite{shi2018,shi2022}.
In the same Al and Cr composition ranges, we compared the phase stability of L1$_2$ FCC structures versus B2 BCC structures using cohesive energies calculated by the trained MTP, following a similar procedure describe in the Section 3.1.
The single phase stability in Figure~\ref{fig:phase-stability} suggests that L1$_2$ FCC structures are more favorable at low Al and Cr compositions whereas B$_2$ BCC structures are more stable at high Al and Cr compositions.
While Figure~\ref{fig:phase-stability} shows the trend of single-phase stability, for a given composition one should always first consider the possibility of forming single phase structures.
In other words, we can use the results in Figure~\ref{fig:phase-stability} only if a single-phase structure tends to form.

\begin{figure}
\centering
\includegraphics[width=5.5in]{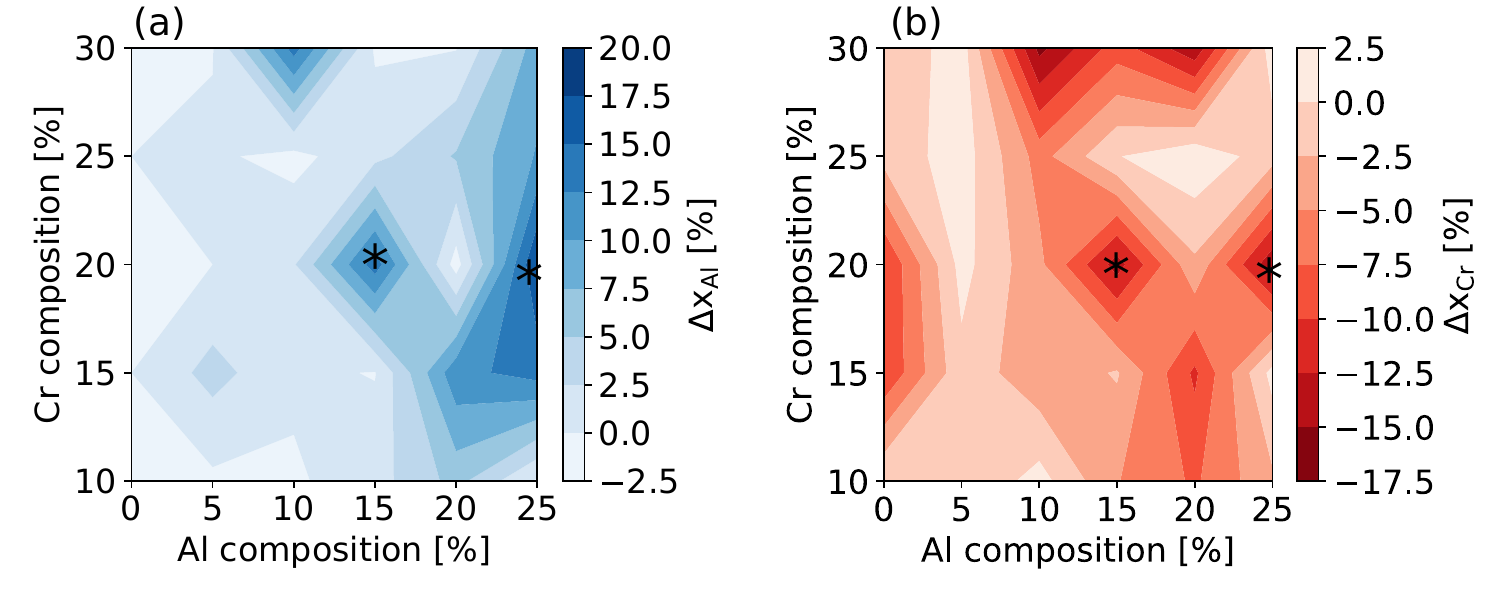}
\caption{Surface segregation of Al and Cr for different Al and Cr compositions for FCC(111) surfaces of Al$_x$Cr$_y$(FeCoNi)$_{100-x-y}$.}
\label{fig:surface-segregation}
\end{figure}

Figure~\ref{fig:results}(b) shows PBR$_{\rm{Cr}}$ for different Al and Cr compositions.
Likewise, a major dependence on Al compositions and a minor dependence on Cr compositions are identified.
Higher PBR$_{\rm{Cr}}$ can be found at higher Al and Cr compositions, and PBR$_{\rm{Cr}}$ values over the composition space studied spans in a range between 2.00 and 2.18, close to the PBR$_{\rm{Cr}}$ of pure Cr (2.04).
PBR values can be used to estimate the average stresses in oxide passive films.
The analysis by Bernstein~\cite{bernstein1987} and Huntz et al.~\cite{huntz1995} suggests that the average stresses are given by $\epsilon = \omega [(PBR)^{1/3}-1]$ where $\omega$ is a correction scaling factor around 0.18.
Using this simple analysis, oxide stresses across over compositions studied have a negligible difference of around 0.5\%.
In contrast, the surface energies exhibit larger variations over compositions, ranging from 1.92 to 2.24 J/m$^2$.
High surface energies are concentrated at the regions with high Al and Cr compositions, while low surface energies are found with Cr contents around 18\% and with either low or high Al contents.
Besides, we can relate the surface energies to electrochemical dissolution rate using Eq.~\ref{eq:rate}.
Using an atomic density of $1 \times 10^{19}~\textrm{atoms/m}^2$ and a transition coefficient $\alpha$ of 1/2, electrochemical dissolution rates over the studied composition ranges vary by 50 folds.
To understand surface energy dependency on Al and Cr compositions, we studied the surface segregation of Al and Cr for each composition, and the results are depicted in Figure~\ref{fig:surface-segregation}. The surface segregation for an element $\rm{M}$ is defined as:

\begin{equation}\label{eq:segreation}
\Delta x_{\rm{M}} = c_{\rm{M}}^{\mathrm{surface}} - c_{\rm{M}}^0
\end{equation}

\noindent
where $c_{M}^{\mathrm{surface}}$ is the surface composition of metal M and $c_{M}^0$ is the composition given by the chemical formula.
It is thus inferred that the low surface energies in the middle and bottom-right regions of Figure~\ref{fig:results}(c) originate from Al segregation and Cr depletion, as marked by `*' in Figure~\ref{fig:surface-segregation} since a Al FCC(111) surface has a much lower surface energy (0.77 J/m$^2$) compared to a Cr FCC(111) surface (2.61 J/m$^2$), as shown in Table~\ref{tab:mtp-val}.
The Cr depletion on the surface lends itself difficult to the formation of \ce{Cr2O3} passive films.
In summary, highly corrosion-resistant AlCrFeCoNi alloys can potentially be found with low Al contents and around 18\% Cr contents because the alloy with AlCrFeCoNi alloys with these compositions tend to form single-phase alloys and to exhibit lower surface energies.
The identified Al compositions are consistent with experimental measurement~\cite{shi2018,shi2020,shi2022}.

\section{Conclusion and outlook}

A machine learning framework was proposed and developed to accelerate the discovery of corrosion-resistant high-entropy alloys.
We demonstrated that the proposed framework can provide an accurate evaluation of relative corrosion resistance for a wide range of compositions for high-entropy alloys.
The physics-informed framework consists of two machine learning approaches.
One approach uses experimental data to train random forest classifier for predictions of single phase formability.
The other approach uses first-principles data to develop robust machine learning potentials, allowing for fast downstream simulations to obtain corrosion metrics such as Pilling-Bedworth ratio and surface energy.
Current computational methods to understand corrosion performance of alloys mostly use pure statistical fitting or first-principles calculations.
Unlike statistical fitting, the random forest classifier encodes meaningful physical knowledge into the feature engineering process.
In comparison with first-principles calculations, the machine learning potentials can significantly mitigate the computational overhead of massive first-principles calculations.
This framework was tested on a specific class of high-entropy alloys AlCrFeCoNi.
The AlCrFeCoNi compositions were sampled by varying the Al and Cr compositions while enforcing the remaining Fe, Co and Ni compositions to be almost identical.
The three corrosion metrics were evaluated on those sampled compositions, based on which the desired compositions for corrosion protection were identified.
We found that low Al compositions and around 18\% Cr compositions tend to form corrosion-resistant alloys, in satisfactory agreement with experimental observations.
Although additional corrosion descriptors, such as cohesive energy, and adsorption energy of oxygen and chloride, may be needed to provide a more comprehensive description of corrosion performance, the three simple corrosion metrics used in this work have proved to be effective in narrowing down the composition space for further selection.

Our scheme is not limited to AlCrFeCoNi high-entropy alloys and corrosion properties.
The methodology can be easily adapted for other material applications where the relationship of chemical compositions with properties is sought after and where ML accelerated molecular simulations are indispensable for high-throughput screening of material candidates.
For instance, ductility of alloys can be evaluated by stacking fault energies which can be calculated by using MLPs, and hardness can be estimated using machine learning regression on experimental dataset~\cite{hu2021,yang2022}.
One should note that most state-of-the-art machine learning potentials use element-specific features, which limit the transferability of MLPs.
In other words, MLPs trained on certain elements cannot be applied to elements not existing in the training data.
Moreover, a large amount of training structures are required to build reliable ML models for high-entropy systems.
Developing a robust element-agnostic featurization method, and reducing numbers of representative images and features are promising future directions.
Element-agnostic featurization methods are just emerging in recent years and needs further development.
For example, there are methods using multipole expansions of the electron density around atoms~\cite{lei2022} or graph representation of materials~\cite{chen2022-2}.
Current image and feature selection methods use linear correlations in the feature space~\cite{imbalzano2018}.
More advanced methods may require understand the complex non-linear mapping between features and outputs (e.g. energy and forces).

\subsection*{Declaration of competing interest}

The authors declare that they have no known competing financial interests or personal relationships that could have appeared to influence the work reported in this paper.

\subsection*{Acknowledgments}

This work was completed in part using the Discovery cluster, supported by the Research Computing team at Northeastern University.
This work is partially supported by The Experiential AI Institute, Roux Institute, and the Alfond Foundation at Northeastern University.
We thank Dr. Liang Qi (University of Michigan) for providing valuable technical feedback to our poster presented at ICME2023 conference in Orlando, Florida, USA.

\subsection*{Supporting information}
Standard enthalpy of formation for oxides of certain metals, Computational settings of DFT calculations, Details of moment tensor potential (MTP), Random forest classifier for single phase formability, Fast sampling of configurations using the embedded atom method (EAM), First-principles data for MTP, MTP-enabled simulations for corrosion metrics, and Supporting results are included in the supporting information.



\subsection*{Data and code availability}
Code and data to reproduce the findings can be found on github: \href{https://github.com/cengc13/ml-hea-corrosion-code-data}{https://github.com/cengc13/ml-hea-corrosion-code-data}.





\bibliography{references}
\bibliographystyle{bibstyle}

\end{document}